\documentclass[prl,twocolumn,superscriptaddress]{revtex4-1}

\usepackage{graphicx}
\newcommand{\bra}{\langle}
\newcommand{\ket}{\rangle}
\newcommand{\up}{\uparrow}
\newcommand{\dn}{\downarrow}
\newcommand{\bbra}{\langle\!\langle}
\newcommand{\kket}{\rangle\!\rangle}

\usepackage{color}

\begin{document}

\title{Long range dynamical coupling between magnetic adatoms mediated by a 2D topological insulator}

\author{M. Costa}
\affiliation{Brazilian Nanotechnology National Laboratory (LNNano), CNPEM, 13083-970 Campinas, Brazil.}
\author{M. Buongiorno Nardelli}
\affiliation{Department of Physics and Department of Chemistry, University of North Texas, Denton TX, USA.}
\affiliation{Center for Materials Genomics, Duke University, Durham, NC 27708, USA.}
\author{A. Fazzio}
\affiliation{Brazilian Nanotechnology National Laboratory (LNNano), CNPEM, 13083-970 Campinas, Brazil.}
\author{A. T. Costa}
\affiliation{Instituto de Telecomunica\c c\~oes, Physics of Information and Quantum Technologies Group, Lisbon, Portugal.}
\affiliation{Brazilian Nanotechnology National Laboratory (LNNano), CNPEM, 13083-970 Campinas, Brazil.}
\affiliation{Instituto de F\'isica, Universidade Federal Fluminense, 24210-346 Niter\'oi, RJ, Brazil.}
\email{antoniocosta@id.uff.br}

%\pacs{75.78.-n,85.75.-d,73.43.-f}

\begin{abstract}
We study the spin excitation spectra and the 
dynamical exchange coupling between iron adatoms on a Bi bilayer 
nanoribbon. We show that the topological character of the edge 
states is preserved in the presence of the magnetic adatoms. 
Nevertheless, they couple significantly to the edge spin
currents, as witnessed by the large and long-ranged dynamical 
coupling we obtain in our calculations. The large 
effective magnetocrystalline anisotropy of the magnetic adatoms 
combined with the transport properties of the topologically protected 
edge states make this system a strong candidate for 
implementation of spintronics devices and quantum information 
and/or computation protocols.
\end{abstract}
 
\maketitle

\section{Introduction}

Topological quantum matter has been recognized as a 
fundamental concept in physics as well as a huge promise 
for future quantum technologies~\cite{NobelLectureRMP_TQM,RevModPhys.82.3045,RevModPhys.83.1057}. 
The peculiar properties of topologically protected edge states 
(TPES) have been hailed as a ``Holy Grail'' for quantum information 
processing~\cite{QI1}. Heterostructures combining topological insulators
and other ``non-trivial'' matter, such as superconductors, hold
promises for even more intriguing states, such as Majorana 
fermions~\cite{MajoranaTIKanePRL2008,ReviewMajoranaBeenacker}.

Magnetic systems of nanoscopic dimensions have also been considered as
candidates for implementation of quantum technologies. Proposals
range from controlling electronic spins in quantum dots~\cite{FabianQDHalfAdder}
to manipulate individual atomic magnetizations on surfaces~\cite{khajetooriansLogicGate}. 

The idea of depositing magnetic atoms on two-dimensional topological 
insulators has particularly enticing aspects. From the fundamental
point of view, it is intriguing to enquire about the strength and
nature of the coupling between magnetic units and the topologically
protected edge states. From the technological side, functionalization
with magnetic adatoms may be an efficient route to tap into the
attractive properties of TPES~\cite{adatom-1,r-wu}. As an added bonus, the high spin-orbit
coupling strength necessary to produce topologically non-trivial states
may also endow magnetic adatoms with magnetocrystalline anisotropy large
enough to stabilize the magnetization direction against quantum and
thermal fluctuations. Two general questions immediately come to
mind: How does the presence of magnetic impurities affect topological 
protection, considering it breaks time-reversal symmetry? and How does the topological nature of the substrate affect 
the properties of the magnetic unities and its interactions? To try and answer these questions
is the main motivation of this work. 

The study of spin excitations provide a prolific source of information
on the properties of magnetic systems. It can reveal the nature and magnitudes
of individual contributions to the system's energy (such as exchange, anisotropy, etc.) 
as well as relaxation and absorption properties. Spin excitations in individual 
magnetic adatoms have been probed experimentally with inelastic tunneling 
spectroscopy~\cite{Heinrich:2004:singleatomspinflip,antc:2011:FeCuSTS},
revealing the peculiarities of the magnetization dynamics at the atomic
scale. Theoretical knowledge of the spin excitation spectra of such 
systems has been instrumental in the design and interpretation of those 
experiments.

\begin{figure}
\includegraphics[width=0.8\columnwidth,angle=-90]{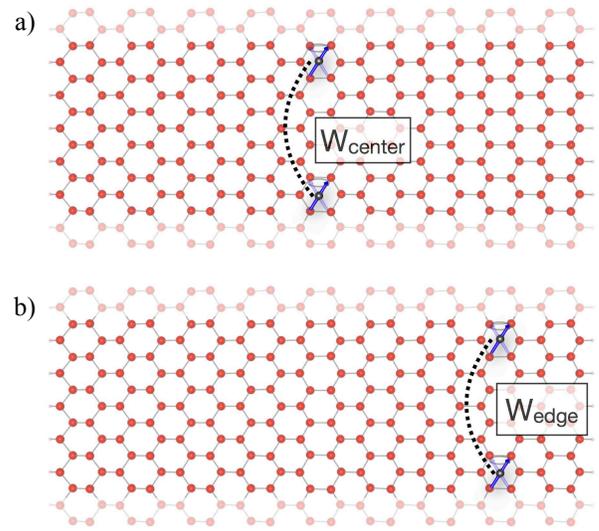}
\caption{Schematic picture of the two magnetic impurities configuration,  at the (a) center and (b) edge of the nanoribbon. The nanoribbon is wide enough to minimize the presence of the topological edge states in its middle region (center). There is a remarkably difference in 
their interaction ($W$) at the two considered positions.}
\label{schematic}
\end{figure}

In this letter we investigate theoretically, employing a 
combination of combination of \textit{ab initio} DFT calculations, 
realistic tight-binding models and many-body techniques, the ground 
state and magnetic excitations of Fe adatoms on a Bi bilayer nanoribbon.
The Bi bilayer is a promising TI with a 500 meV gap~\cite{bi-1,bi-2,bi-3,bi-4,bi-5,bi-6}, which is originated from the
large Bi SOC. We consider isolated adatoms and pairs of adatoms deposited on
an otherwise pristine ribbon. The ribbon's width is large
enough to guarantee the existence of topologically protected edge
states~\cite{bi-4}. We investigate the indirect exchange coupling (sometimes
dubbed RKKY coupling) through the substrate and show that, for the 
smallest distances considered, it is essentially zero. We focus our 
attention on the dynamical coupling that appears as a result of the 
exchange of pure spin currents between the out-of-equilibrium
magnetizations of the Fe adatoms. We show that it is very long
ranged as a result of the peculiar properties of the TPES.

\section{Methodology}

We use a multi-orbital tight-binding Hamiltonian to describe the 
electronic structure of the decorated nanoribbon. The full 
hamiltonian can be written as the sum of band energy $H_0$, 
effective intra-atomic Coulomb repulsion $H_I$ and the atomic spin-orbit
coupling $H_{SOC}$. The band energy $H_0$ is given by,
\begin{equation}
H_0=\sum_{ll'}\sum_{\mu\nu}\sum_{\sigma}T_{ll'}^{\mu\nu}
c^\dagger_{l\mu\sigma}c_{l'\nu\sigma}.
\label{hopping}
\end{equation}
where $T_{ll'}^{\mu\nu}$ is the hopping matrix, $\sigma$ is a spin index, 
$l,l'$ are atomic site indices and $\mu,\nu$ are atomic orbital indices. 
The hopping matrix is obtained directly from a DFT calculation
using the pseudo atomic orbital projection method~\cite{PAO1,PAO2,PAO3,PAO4,PAO5}. 
The method consists in projecting the Hilbert space spanned by the plane waves 
onto a compact subspace composed of the pseudo atomic orbitals (PAO). These PAOs 
functions are naturally built into the pseudo potential used in the DFT 
calculation.

The atomic spin-orbit coupling (SOC) is introduced in a effective approximation,
\begin{equation}
H_\mathrm{SOC}=\sum_l\sum_{\mu\nu}\sum_{\sigma\sigma'}\xi^{\mu\nu}_l
\bra l\mu\sigma|\vec{L}\cdot\vec{S}|l\nu\sigma'\ket c^\dagger_{l\mu\sigma}c_{l\nu\sigma'}
\label{LdotS}
\end{equation}
where $\vec{L}$ and $\vec{S}$ are the orbital and spin angular momentum operators, 
respectively. The SOC strength is given by $\xi^{\mu\nu}_l$ and is also obtained 
from the DFT calculations.

The magnetism of the Fe adatoms is driven by the effective intra-atomic
Coulomb repulsion
\begin{equation}
H_I=\sum_l\sum_{\mu\nu\mu'\nu'}\sum_{\sigma\sigma'}U_l^{\mu\nu\mu'\nu'}
c^\dagger_{l\mu\sigma}c^\dagger_{l\nu\sigma'}c_{l\nu'\sigma'}c_{l\mu'\sigma} .
\label{coulomb}
\end{equation}
The spin-polarized ground state is obtained from a self-consistent mean-field
treatment of the effective Coulomb repulsion term, Eq.~(\ref{coulomb}). Both magnitude
and direction of the magnetization at each adatom are found as solutions of the 
self-consistency equations. After a solution is found, a calculation of the spin
excitation spectrum reveals whether it is a stable or unstable solution. In the
latter case, we search for a new solution, usually starting from a high symmetry
configuration. The comparison between the band structures from the DFT and the 
multi-orbital tight-binding Hamiltonian are shown in the supplemental material~\cite{SM} 
along with the procedure to obtain the SOC strength $\xi^{\mu\nu}_l$.

The spin excitation spectrum is obtained from the spin susceptibility
matrix $\chi^{ab}_{\mu\nu}(l,l')$, where $a,b\in\{+,-,\up,\dn\}$~\cite{antc:2010:SOCMethod,FilipePRB2015}. For instance,
if the equilibrium magnetization lies along the $z$ direction, the transverse 
spin excitation spectrum is given by $\chi^{+-}_{\mu\nu}(l,l')$. For a general magnetization vector, however, a more complicated combination of the matrix elements $\chi^{ab}$
will result from applying to it the appropriate rotation transformations.

The spin susceptibility matrix elements are the propagators
\begin{equation}
\chi^{ab}_{\mu\nu}(l,l'|t-t') = -i\theta(t-t')\bbra S^a_{l\mu}(t),S^{b}_{l'\nu}(t')\kket .
\label{definechi}
\end{equation}
Due to the interaction term Eq.~\ref{coulomb}, the equations of motion for $\chi^{ab}$
are coupled to higher order propagators, in an infinite chain. We employ a decoupling 
scheme that is frequently dubbed ``time dependent Hartree-Fock''. It is equivalent to 
the summation of the infinite series for the electron-hole pair ladder diagrams, and leads to
the collective spin excitation modes identified with the poles of $\chi^{ab}$. 
In translation-invariant systems those
modes are the spin waves; when the magnetic entities are confined in space
the spin excitation modes are akin to normal modes of coupled oscillators.
In the presence of spin-orbit coupling the transverse spin excitations are
coupled to longitudinal excitations and charge excitations. This is reflected
in the fact that the equations of motion for all the elements of $\chi^{ab}$
have to be solved simultaneously. The solution has been presented in detail 
in reference~\cite{antc:2010:SOCMethod}. The physical consequences of this 
dynamical spin-charge coupling have been little explored hitherto, but we intend to
divulge the results of our investigations on this matter in future publications.

When considered as a matrix in site indices, the transverse susceptibility
$\chi^{+-}_{ll'}$ yields two very relevant sets of informations: the imaginary
part of its diagonal elements represent the local (atomic site) projection
of the magnon spectral density, or the local magnon density of states. From it
we can extract the magnon's energies and lifetimes. The off-diagonal terms
represent non-local response functions, and inform about the excitation of
a magnon at site $l$ due to the application of a transverse field at site $l'$.
It is important to notice that this ``dynamical coupling'' may exist even 
when $|\vec{R}_l-\vec{R}_{l'}|$ is so large that the (RKKY) indirect exchange
coupling is utterly negligible~\cite{rkky1,rkky2}. Its physical origin is the 
spin current shed
by the spin excitation at $l'$, which flows through the supporting medium
and reaches site $l$, thus imparting a torque on the magnetization there. 
We chose to represent the intensity of the dynamical coupling between sites
$l$ and $l'$ as the amplitude of the magnetization precession at site $l$
due to a transverse, circularly polarized magnetic field applied at site $l'$.

\section{Results and discussion}
\begin{figure}
\includegraphics[width=0.9\columnwidth,angle=-90]{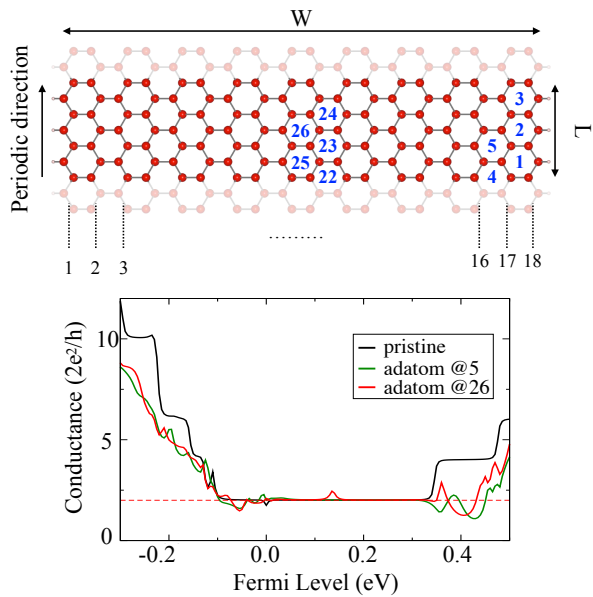}
\caption{Top panel shows the Bi zigzag nanoribbon structure with 
18 zigzag units and a width (W) of 65~\AA. The length (L) is equal 
to 13.05~\AA~along the periodic direction. The adsorption sites are
labeled, from right to left, as showed by the blue numbers. 
In bottom panel the pristine, adatom @26 and adatom @5 conductance is 
showed.}
\label{conductance}
\end{figure}

We begin by showing that the topologically protected edge states 
(TPES) of the Bi bilayer nanoribbon are negligibly perturbed by 
the presence of the adatoms. In figure~\ref{conductance}, bottom panel, we 
compare the conductance of the pristine nanoribbon with that of 
the ribbons decorated with a single Fe adatom. It is clear that 
the conductance of the TPES is preserved over most of the bulk 
gap for all adatoms positions considered. The behavior of the
edge spin currents as a function of energy are modified by the 
adatoms, but their overall order of magnitude is preserved. 

The magnetic moments of the adatoms have been determined 
selfconsistently, both in size and direction. For a single adatom 
at the center of the ribbon the magnetic moment is mostly 
perpendicular to the plane of the ribbon, with a small in-plane 
component perpendicular to the ribbon's length. For the adatom 
close to the edge the magnetic moment becomes almost parallel to 
the plane of the ribbon.

% 1 meV = 2.418 x 10^{11} Hz

\begin{figure}
\includegraphics[width=0.9\columnwidth]{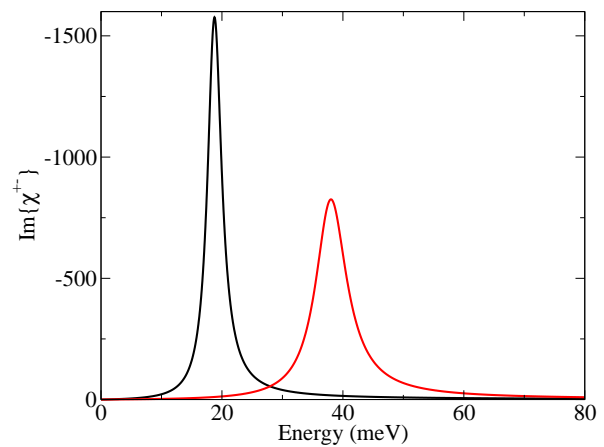}
\caption{Spectral densities for transverse spin excitations
of single adatoms adsorbed to the Bi bilayer nanoribbon at
two distinct sites: close to one edge (@5, black line)
and at the center of the ribbon (@26, red line).}
\label{chi_isolated}
\end{figure}

The spectral densities for transverse spin excitations for single 
adatoms at different positions across the ribbon are shown in 
figure~\ref{chi_isolated}. From those spectral densities we can 
extract an effective magnetocrystalline anisotropy energy 
(eMAE)~\footnote[1]{In linear response theory the system is assumed to
be arbitrarily close to equilibrium at all times. This means that the
excitation energy is determined by the ground state electronic 
structure, and may differ from the MAE calculated via total energy
differences.}
and the damping rate (or equivalently, the 
lifetime) of the transverse spin excitations. For the adatom 
at the center of the ribbon the eMAE is 38~meV, which corresponds
to a precession frequency of approximately 9.2~THz. The eMAE for 
the adatom close to the edge is 19~meV ($\nu_0\approx 4.6$~THz). 
Although the nominal excitation lifetime for the adatom close to 
the edge is larger, the quality factors for both positions are 
very similar ($Q=12$ for the center and  $Q=13$ close to 
the edge). Notice that the MAE we obtained is completely due
to the huge spin-orbit coupling strength of Bi 
($\xi_\mathrm{Bi}=1.475$~eV), since the tiny SOC strength of 
Fe ($\xi_\mathrm{Fe}=0.08$~eV)  was not even included 
in our calculations.

It is interesting to notice that transverse spin excitations of 
adatoms on insulating surfaces usually have much longer 
lifetimes than those we obtained.~\cite{Heinrich:2004:singleatomspinflip}
The large damping is certainly due to the huge SOC strength of Bi.
However, we speculate that the spin angular momentum associated 
with the spin excitation has first to penetrate into the Bi 
substrate in order to be then transferred to the orbital degrees 
of freedom and ultimately be dissipated by the lattice vibrations.
The fact that spin currents can easily flow into a non-magnetic 
insulating substrate may seem somewhat counterintuitive, but in 
what follows we will show that this is indeed the case.

\begin{figure}
\begin{center}
\includegraphics[width=0.95\columnwidth]{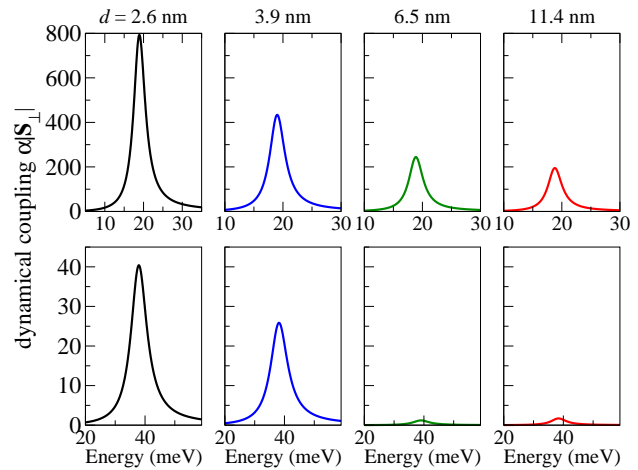}
\end{center}
\caption{Dynamical coupling as a function of excitation frequency
for selected inter-adatoms distances. The top panel corresponds 
to the adatoms adsorbed close to one of the ribbon's edges (@5). 
The bottom panel corresponds to the adatoms adsorbed to the 
center of the ribbon (@26).}
\label{Sxy_x_W}
\end{figure}

\begin{figure}
\begin{center}
\includegraphics[width=0.95\columnwidth]{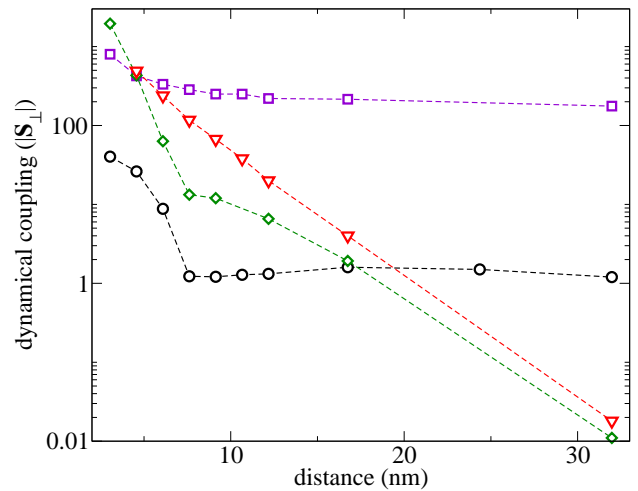}
\end{center}
\caption{Dynamical coupling as a function of the inter-adatom
distance at resonance. Circles and squares correspond to the 
adatoms adsorbed to the center and close to the edge of the 
ribbon, respectively. The values depicted using triangles and 
diamonds were obtained by artificially suppressing the strength 
of spin-orbit coupling in Bi to 0.8~eV, such that no topologically 
protected edge states exist in the nanoribbon.}
\label{dyn_coup}
\end{figure}

We now turn to the spin excitations of pairs of Fe adatoms. 
We are mainly interested in the behavior of the transverse spin 
excitation spectra as the distance between the adatoms is varied. 
As we will show, for almost all distances considered there is no 
Heisenberg-like indirect exchange coupling between the adatoms. 
However, a large dynamical coupling appears as soon as the  
magnetic moments are driven away from equilibrium in a 
time-dependent manner. We choose as a measure of the dynamical 
coupling the norm of the transverse magnetization induced at 
adatom 2 due to a transverse field applied to adatom 1.
This is proportional to the non-local susceptibility 
$\chi^{+-}_{12}$, as detailed in reference~\cite{antc:2008:DynamicCouplingToyModel}. 
In figure~\ref{Sxy_x_W} we show the dynamical coupling as a 
function of energy for various distances between the adatoms. 
In each case the adatoms are placed in the nanoribbon on the
same position relative to the edge. In the top panel the 
adatoms are close to the edge and in the bottom panel they are 
adsorbed at the center of the nanoribbon. It is clear that the 
dynamical coupling decreases abruptly as a function of distance 
for the adatoms at the central sites, but only slightly
for the adatoms close to the edge. This is an indication that the 
topologically protected edge states have an essential role in 
mediating the dynamical coupling between the adatoms. To highlight
the importance of the topologically protected edge states as 
mediators of the dynamical coupling, we reduced by hand the
spin-orbit coupling in Bi to 0.8~eV and repeated the 
calculations. For this value of the SOC strength the Bi bilayer
nanoribbon is a non-topological insulator and there are no
TPES. In figure~\ref{dyn_coup} we compare the behavior of the 
dynamical coupling as a function of the distance between the 
adatoms in the two cases, ``real'' Bi, and Bi with an 
artificially reduced SOC strength. For both the adatom close to 
the edge and at the center of the ribbon, the dynamical coupling 
decays exponentially with the distance for the reduced SOC case. 
This is markedly different from the behavior of the real system. 
There, only a slight, polynomial decay is observed when the 
adatoms are close to the edge. For the adatoms at the center the 
decay is faster, although definitely not exponential, and the 
dynamic coupling reaches a plateau for distances larger than 
about 8~nm. We interpret these results as evidence that the 
topologically protected edge states play a crucial role in 
carrying the spin currents responsible for the dynamical coupling
between the adatoms' magnetic moments. There is a curious 
interplay here, because the strong SOC of Bi would be expected 
to scramble spin information rapidly for spin currents flowing 
through a non-topological material made of Bi. In the language 
commonly used in the spin transport community, Bi materials would
be considered almost ideal spin sinks. The existence of
topologically protected edge states, however, render Bi bilayer
an almost ideal spin current conduit, enabling spin information
to be transported through large distances with very small losses.

\section{Concluding remarks}
We used a combination of DFT calculations and multi-orbital 
tight-binding hamiltonians to study the collective spin 
excitations of magnetic adatoms on a two-dimensional topological 
insulator. We have shown that the large spin-orbit coupling in 
Bi induces unusually large effective magnetocrystalline anisotropy energies 
in the Fe adatoms, which result in spin excitation frequencies within the 
terahertz range. Most importantly, we have shown that the 
topologically protected edge states of the Bi bilayer nanoribbon
mediate a long range dynamical coupling between the adatoms' 
magnetic moments. This coupling is a consequence of the
almost unperturbed flow of spin currents between two distant
adatoms. We believe these findings indicate an effective way to couple 
external probes to topologically protected edge states. 
They also highlight the unusual and potentially very
useful spin transport properties of two-dimensional topological 
insulators. Moreover, the findings we reported show unequivocally 
that there is much to be learned by building hybrid structures from 
topological insulators and magnetic materials.

\section{acknowledgement}
This work was supported by the Brazilian agencies FAPESP, through the Grant TEM\'{ATICO} (2017/02327-2), 
and CNPq. We would like to acknowledge computing time provided by Laborat\'orio de 
Computa\c{c}\~ao  Cien\'ifica Avancada  (Universidade de S\~ao Paulo). 
MBN acknowledges support by DOD-ONR (N00014-13-1-0635, N00014-11-1-0136,N00014-15-1-2863), 
the High Performance Computing Center at the University of North Texas, and the 
Texas Advanced Computing Center at the University of Texas, Austin.
ATC wishes to acknowledge partial financial support from Funda\c c\~ao para
a Ci\^encia e a Tecnologia (Portugal), namely through programmes PTDC/POPH/POCH and 
projects UID/EEA/50008/2013, IT/QuNet, partially funded by EU FEDER, from the QuantERA project TheBlinQC, and from the JTF project NQuN (ID 60478), and FAPESP (Brazil). ATC is also indebted 
to R. Bechara Muniz and F. S. M. Guimar\~aes for enlightening discussions.

\bibliography{Fe_on_Bi_letter}

%merlin.mbs apsrev4-1.bst 2010-07-25 4.21a (PWD, AO, DPC) hacked
%Control: key (0)
%Control: author (72) initials jnrlst
%Control: editor formatted (1) identically to author
%Control: production of article title (-1) disabled
%Control: page (0) single
%Control: year (1) truncated
%Control: production of eprint (0) enabled
\begin{thebibliography}{30}%
\makeatletter
\providecommand \@ifxundefined [1]{%
 \@ifx{#1\undefined}
}%
\providecommand \@ifnum [1]{%
 \ifnum #1\expandafter \@firstoftwo
 \else \expandafter \@secondoftwo
 \fi
}%
\providecommand \@ifx [1]{%
 \ifx #1\expandafter \@firstoftwo
 \else \expandafter \@secondoftwo
 \fi
}%
\providecommand \natexlab [1]{#1}%
\providecommand \enquote  [1]{``#1''}%
\providecommand \bibnamefont  [1]{#1}%
\providecommand \bibfnamefont [1]{#1}%
\providecommand \citenamefont [1]{#1}%
\providecommand \href@noop [0]{\@secondoftwo}%
\providecommand \href [0]{\begingroup \@sanitize@url \@href}%
\providecommand \@href[1]{\@@startlink{#1}\@@href}%
\providecommand \@@href[1]{\endgroup#1\@@endlink}%
\providecommand \@sanitize@url [0]{\catcode `\\12\catcode `\$12\catcode
  `\&12\catcode `\#12\catcode `\^12\catcode `\_12\catcode `\%12\relax}%
\providecommand \@@startlink[1]{}%
\providecommand \@@endlink[0]{}%
\providecommand \url  [0]{\begingroup\@sanitize@url \@url }%
\providecommand \@url [1]{\endgroup\@href {#1}{\urlprefix }}%
\providecommand \urlprefix  [0]{URL }%
\providecommand \Eprint [0]{\href }%
\providecommand \doibase [0]{http://dx.doi.org/}%
\providecommand \selectlanguage [0]{\@gobble}%
\providecommand \bibinfo  [0]{\@secondoftwo}%
\providecommand \bibfield  [0]{\@secondoftwo}%
\providecommand \translation [1]{[#1]}%
\providecommand \BibitemOpen [0]{}%
\providecommand \bibitemStop [0]{}%
\providecommand \bibitemNoStop [0]{.\EOS\space}%
\providecommand \EOS [0]{\spacefactor3000\relax}%
\providecommand \BibitemShut  [1]{\csname bibitem#1\endcsname}%
\let\auto@bib@innerbib\@empty
%</preamble>
\bibitem [{\citenamefont {Haldane}(2017)}]{NobelLectureRMP_TQM}%
  \BibitemOpen
  \bibfield  {author} {\bibinfo {author} {\bibfnamefont {F.~D.~M.}\
  \bibnamefont {Haldane}},\ }\href {\doibase 10.1103/RevModPhys.89.040502}
  {\bibfield  {journal} {\bibinfo  {journal} {Rev. Mod. Phys.}\ }\textbf
  {\bibinfo {volume} {89}},\ \bibinfo {pages} {040502} (\bibinfo {year}
  {2017})}\BibitemShut {NoStop}%
\bibitem [{\citenamefont {Hasan}\ and\ \citenamefont
  {Kane}(2010)}]{RevModPhys.82.3045}%
  \BibitemOpen
  \bibfield  {author} {\bibinfo {author} {\bibfnamefont {M.~Z.}\ \bibnamefont
  {Hasan}}\ and\ \bibinfo {author} {\bibfnamefont {C.~L.}\ \bibnamefont
  {Kane}},\ }\href {\doibase 10.1103/RevModPhys.82.3045} {\bibfield  {journal}
  {\bibinfo  {journal} {Rev. Mod. Phys.}\ }\textbf {\bibinfo {volume} {82}},\
  \bibinfo {pages} {3045} (\bibinfo {year} {2010})}\BibitemShut {NoStop}%
\bibitem [{\citenamefont {Qi}\ and\ \citenamefont
  {Zhang}(2011)}]{RevModPhys.83.1057}%
  \BibitemOpen
  \bibfield  {author} {\bibinfo {author} {\bibfnamefont {X.-L.}\ \bibnamefont
  {Qi}}\ and\ \bibinfo {author} {\bibfnamefont {S.-C.}\ \bibnamefont {Zhang}},\
  }\href {\doibase 10.1103/RevModPhys.83.1057} {\bibfield  {journal} {\bibinfo
  {journal} {Rev. Mod. Phys.}\ }\textbf {\bibinfo {volume} {83}},\ \bibinfo
  {pages} {1057} (\bibinfo {year} {2011})}\BibitemShut {NoStop}%
\bibitem [{\citenamefont {Ferreira}\ and\ \citenamefont {Loss}(2013)}]{QI1}%
  \BibitemOpen
  \bibfield  {author} {\bibinfo {author} {\bibfnamefont {G.~J.}\ \bibnamefont
  {Ferreira}}\ and\ \bibinfo {author} {\bibfnamefont {D.}~\bibnamefont
  {Loss}},\ }\href {\doibase 10.1103/PhysRevLett.111.106802} {\bibfield
  {journal} {\bibinfo  {journal} {Phys. Rev. Lett.}\ }\textbf {\bibinfo
  {volume} {111}},\ \bibinfo {pages} {106802} (\bibinfo {year}
  {2013})}\BibitemShut {NoStop}%
\bibitem [{\citenamefont {Fu}\ and\ \citenamefont
  {Kane}(2008)}]{MajoranaTIKanePRL2008}%
  \BibitemOpen
  \bibfield  {author} {\bibinfo {author} {\bibfnamefont {L.}~\bibnamefont
  {Fu}}\ and\ \bibinfo {author} {\bibfnamefont {C.~L.}\ \bibnamefont {Kane}},\
  }\href {\doibase 10.1103/PhysRevLett.100.096407} {\bibfield  {journal}
  {\bibinfo  {journal} {Phys. Rev. Lett.}\ }\textbf {\bibinfo {volume} {100}},\
  \bibinfo {pages} {096407} (\bibinfo {year} {2008})}\BibitemShut {NoStop}%
\bibitem [{\citenamefont {Beenakker}(2013)}]{ReviewMajoranaBeenacker}%
  \BibitemOpen
  \bibfield  {author} {\bibinfo {author} {\bibfnamefont {C.}~\bibnamefont
  {Beenakker}},\ }\href {\doibase 10.1146/annurev-conmatphys-030212-184337}
  {\bibfield  {journal} {\bibinfo  {journal} {Annual Review of Condensed Matter
  Physics}\ }\textbf {\bibinfo {volume} {4}},\ \bibinfo {pages} {113} (\bibinfo
  {year} {2013})},\ \Eprint
  {http://arxiv.org/abs/https://doi.org/10.1146/annurev-conmatphys-030212-184337}
  {https://doi.org/10.1146/annurev-conmatphys-030212-184337} \BibitemShut
  {NoStop}%
\bibitem [{\citenamefont {Pfeffer}\ \emph {et~al.}(2016)\citenamefont
  {Pfeffer}, \citenamefont {Hartmann}, \citenamefont {Neri}, \citenamefont
  {Schade}, \citenamefont {Emmerling}, \citenamefont {Kamp}, \citenamefont
  {Gammaitoni}, \citenamefont {Höfling},\ and\ \citenamefont
  {Worschech}}]{FabianQDHalfAdder}%
  \BibitemOpen
  \bibfield  {author} {\bibinfo {author} {\bibfnamefont {P.}~\bibnamefont
  {Pfeffer}}, \bibinfo {author} {\bibfnamefont {F.}~\bibnamefont {Hartmann}},
  \bibinfo {author} {\bibfnamefont {I.}~\bibnamefont {Neri}}, \bibinfo {author}
  {\bibfnamefont {A.}~\bibnamefont {Schade}}, \bibinfo {author} {\bibfnamefont
  {M.}~\bibnamefont {Emmerling}}, \bibinfo {author} {\bibfnamefont
  {M.}~\bibnamefont {Kamp}}, \bibinfo {author} {\bibfnamefont {L.}~\bibnamefont
  {Gammaitoni}}, \bibinfo {author} {\bibfnamefont {S.}~\bibnamefont
  {Höfling}}, \ and\ \bibinfo {author} {\bibfnamefont {L.}~\bibnamefont
  {Worschech}},\ }\href {http://stacks.iop.org/0957-4484/27/i=21/a=215201}
  {\bibfield  {journal} {\bibinfo  {journal} {Nanotechnology}\ }\textbf
  {\bibinfo {volume} {27}},\ \bibinfo {pages} {215201} (\bibinfo {year}
  {2016})}\BibitemShut {NoStop}%
\bibitem [{\citenamefont {Khajetoorians}\ \emph
  {et~al.}(2011{\natexlab{a}})\citenamefont {Khajetoorians}, \citenamefont
  {Wiebe}, \citenamefont {Chilian},\ and\ \citenamefont
  {Wiesendanger}}]{khajetooriansLogicGate}%
  \BibitemOpen
  \bibfield  {author} {\bibinfo {author} {\bibfnamefont {A.~A.}\ \bibnamefont
  {Khajetoorians}}, \bibinfo {author} {\bibfnamefont {J.}~\bibnamefont
  {Wiebe}}, \bibinfo {author} {\bibfnamefont {B.}~\bibnamefont {Chilian}}, \
  and\ \bibinfo {author} {\bibfnamefont {R.}~\bibnamefont {Wiesendanger}},\
  }\href {\doibase 10.1126/science.1201725} {\bibfield  {journal} {\bibinfo
  {journal} {Science}\ }\textbf {\bibinfo {volume} {332}},\ \bibinfo {pages}
  {1062} (\bibinfo {year} {2011}{\natexlab{a}})}\BibitemShut {NoStop}%
\bibitem [{\citenamefont {Honolka}\ \emph {et~al.}(2012)\citenamefont
  {Honolka}, \citenamefont {Khajetoorians}, \citenamefont {Sessi},
  \citenamefont {Wehling}, \citenamefont {Stepanow}, \citenamefont {Mi},
  \citenamefont {Iversen}, \citenamefont {Schlenk}, \citenamefont {Wiebe},
  \citenamefont {Brookes}, \citenamefont {Lichtenstein}, \citenamefont
  {Hofmann}, \citenamefont {Kern},\ and\ \citenamefont
  {Wiesendanger}}]{adatom-1}%
  \BibitemOpen
  \bibfield  {author} {\bibinfo {author} {\bibfnamefont {J.}~\bibnamefont
  {Honolka}}, \bibinfo {author} {\bibfnamefont {A.~A.}\ \bibnamefont
  {Khajetoorians}}, \bibinfo {author} {\bibfnamefont {V.}~\bibnamefont
  {Sessi}}, \bibinfo {author} {\bibfnamefont {T.~O.}\ \bibnamefont {Wehling}},
  \bibinfo {author} {\bibfnamefont {S.}~\bibnamefont {Stepanow}}, \bibinfo
  {author} {\bibfnamefont {J.-L.}\ \bibnamefont {Mi}}, \bibinfo {author}
  {\bibfnamefont {B.~B.}\ \bibnamefont {Iversen}}, \bibinfo {author}
  {\bibfnamefont {T.}~\bibnamefont {Schlenk}}, \bibinfo {author} {\bibfnamefont
  {J.}~\bibnamefont {Wiebe}}, \bibinfo {author} {\bibfnamefont {N.~B.}\
  \bibnamefont {Brookes}}, \bibinfo {author} {\bibfnamefont {A.~I.}\
  \bibnamefont {Lichtenstein}}, \bibinfo {author} {\bibfnamefont
  {P.}~\bibnamefont {Hofmann}}, \bibinfo {author} {\bibfnamefont
  {K.}~\bibnamefont {Kern}}, \ and\ \bibinfo {author} {\bibfnamefont
  {R.}~\bibnamefont {Wiesendanger}},\ }\href {\doibase
  10.1103/PhysRevLett.108.256811} {\bibfield  {journal} {\bibinfo  {journal}
  {Phys. Rev. Lett.}\ }\textbf {\bibinfo {volume} {108}},\ \bibinfo {pages}
  {256811} (\bibinfo {year} {2012})}\BibitemShut {NoStop}%
\bibitem [{\citenamefont {Kim}\ and\ \citenamefont {Wu}(2018)}]{r-wu}%
  \BibitemOpen
  \bibfield  {author} {\bibinfo {author} {\bibfnamefont {J.}~\bibnamefont
  {Kim}}\ and\ \bibinfo {author} {\bibfnamefont {R.}~\bibnamefont {Wu}},\
  }\href {\doibase 10.1103/PhysRevB.97.115151} {\bibfield  {journal} {\bibinfo
  {journal} {Phys. Rev. B}\ }\textbf {\bibinfo {volume} {97}},\ \bibinfo
  {pages} {115151} (\bibinfo {year} {2018})}\BibitemShut {NoStop}%
\bibitem [{\citenamefont {Heinrich}\ \emph {et~al.}(2004)\citenamefont
  {Heinrich}, \citenamefont {Gupta}, \citenamefont {Lutz},\ and\ \citenamefont
  {Eigler}}]{Heinrich:2004:singleatomspinflip}%
  \BibitemOpen
  \bibfield  {author} {\bibinfo {author} {\bibfnamefont {A.~J.}\ \bibnamefont
  {Heinrich}}, \bibinfo {author} {\bibfnamefont {J.~A.}\ \bibnamefont {Gupta}},
  \bibinfo {author} {\bibfnamefont {C.~P.}\ \bibnamefont {Lutz}}, \ and\
  \bibinfo {author} {\bibfnamefont {D.~M.}\ \bibnamefont {Eigler}},\ }\href
  {\doibase 10.1126/science.1101077} {\bibfield  {journal} {\bibinfo  {journal}
  {Science}\ }\textbf {\bibinfo {volume} {306}},\ \bibinfo {pages} {466}
  (\bibinfo {year} {2004})}\BibitemShut {NoStop}%
\bibitem [{\citenamefont {Khajetoorians}\ \emph
  {et~al.}(2011{\natexlab{b}})\citenamefont {Khajetoorians}, \citenamefont
  {Lounis}, \citenamefont {Chilian}, \citenamefont {Costa}, \citenamefont
  {Zhou}, \citenamefont {Mills}, \citenamefont {Wiebe},\ and\ \citenamefont
  {Wiesendanger}}]{antc:2011:FeCuSTS}%
  \BibitemOpen
  \bibfield  {author} {\bibinfo {author} {\bibfnamefont {A.~A.}\ \bibnamefont
  {Khajetoorians}}, \bibinfo {author} {\bibfnamefont {S.}~\bibnamefont
  {Lounis}}, \bibinfo {author} {\bibfnamefont {B.}~\bibnamefont {Chilian}},
  \bibinfo {author} {\bibfnamefont {A.~T.}\ \bibnamefont {Costa}}, \bibinfo
  {author} {\bibfnamefont {L.}~\bibnamefont {Zhou}}, \bibinfo {author}
  {\bibfnamefont {D.~L.}\ \bibnamefont {Mills}}, \bibinfo {author}
  {\bibfnamefont {J.}~\bibnamefont {Wiebe}}, \ and\ \bibinfo {author}
  {\bibfnamefont {R.}~\bibnamefont {Wiesendanger}},\ }\href@noop {} {\bibfield
  {journal} {\bibinfo  {journal} {Phys. Rev. Lett.}\ }\textbf {\bibinfo
  {volume} {106}},\ \bibinfo {pages} {037205} (\bibinfo {year}
  {2011}{\natexlab{b}})}\BibitemShut {NoStop}%
\bibitem [{\citenamefont {Murakami}(2006)}]{bi-1}%
  \BibitemOpen
  \bibfield  {author} {\bibinfo {author} {\bibfnamefont {S.}~\bibnamefont
  {Murakami}},\ }\href {\doibase 10.1103/PhysRevLett.97.236805} {\bibfield
  {journal} {\bibinfo  {journal} {Phys. Rev. Lett.}\ }\textbf {\bibinfo
  {volume} {97}},\ \bibinfo {pages} {236805} (\bibinfo {year}
  {2006})}\BibitemShut {NoStop}%
\bibitem [{\citenamefont {Liu}\ \emph {et~al.}(2011)\citenamefont {Liu},
  \citenamefont {Liu}, \citenamefont {Wu}, \citenamefont {Duan}, \citenamefont
  {Liu},\ and\ \citenamefont {Wu}}]{bi-2}%
  \BibitemOpen
  \bibfield  {author} {\bibinfo {author} {\bibfnamefont {Z.}~\bibnamefont
  {Liu}}, \bibinfo {author} {\bibfnamefont {C.-X.}\ \bibnamefont {Liu}},
  \bibinfo {author} {\bibfnamefont {Y.-S.}\ \bibnamefont {Wu}}, \bibinfo
  {author} {\bibfnamefont {W.-H.}\ \bibnamefont {Duan}}, \bibinfo {author}
  {\bibfnamefont {F.}~\bibnamefont {Liu}}, \ and\ \bibinfo {author}
  {\bibfnamefont {J.}~\bibnamefont {Wu}},\ }\href {\doibase
  10.1103/PhysRevLett.107.136805} {\bibfield  {journal} {\bibinfo  {journal}
  {Phys. Rev. Lett.}\ }\textbf {\bibinfo {volume} {107}},\ \bibinfo {pages}
  {136805} (\bibinfo {year} {2011})}\BibitemShut {NoStop}%
\bibitem [{\citenamefont {Drozdov}\ \emph {et~al.}(2014)\citenamefont
  {Drozdov}, \citenamefont {Alexandradinata}, \citenamefont {Jeon},
  \citenamefont {Nadj-Perge}, \citenamefont {Ji}, \citenamefont {Cava},
  \citenamefont {Andrei~Bernevig},\ and\ \citenamefont {Yazdani}}]{bi-3}%
  \BibitemOpen
  \bibfield  {author} {\bibinfo {author} {\bibfnamefont {I.~K.}\ \bibnamefont
  {Drozdov}}, \bibinfo {author} {\bibfnamefont {A.}~\bibnamefont
  {Alexandradinata}}, \bibinfo {author} {\bibfnamefont {S.}~\bibnamefont
  {Jeon}}, \bibinfo {author} {\bibfnamefont {S.}~\bibnamefont {Nadj-Perge}},
  \bibinfo {author} {\bibfnamefont {H.}~\bibnamefont {Ji}}, \bibinfo {author}
  {\bibfnamefont {R.~J.}\ \bibnamefont {Cava}}, \bibinfo {author}
  {\bibfnamefont {B.}~\bibnamefont {Andrei~Bernevig}}, \ and\ \bibinfo {author}
  {\bibfnamefont {A.}~\bibnamefont {Yazdani}},\ }\href@noop {} {\bibfield
  {journal} {\bibinfo  {journal} {Nature Physics}\ }\textbf {\bibinfo {volume}
  {10}},\ \bibinfo {pages} {664} (\bibinfo {year} {2014})}\BibitemShut
  {NoStop}%
\bibitem [{\citenamefont {Huang}\ \emph {et~al.}(2013)\citenamefont {Huang},
  \citenamefont {Chuang}, \citenamefont {Hsu}, \citenamefont {Liu},
  \citenamefont {Chang}, \citenamefont {Lin},\ and\ \citenamefont
  {Bansil}}]{bi-4}%
  \BibitemOpen
  \bibfield  {author} {\bibinfo {author} {\bibfnamefont {Z.-Q.}\ \bibnamefont
  {Huang}}, \bibinfo {author} {\bibfnamefont {F.-C.}\ \bibnamefont {Chuang}},
  \bibinfo {author} {\bibfnamefont {C.-H.}\ \bibnamefont {Hsu}}, \bibinfo
  {author} {\bibfnamefont {Y.-T.}\ \bibnamefont {Liu}}, \bibinfo {author}
  {\bibfnamefont {H.-R.}\ \bibnamefont {Chang}}, \bibinfo {author}
  {\bibfnamefont {H.}~\bibnamefont {Lin}}, \ and\ \bibinfo {author}
  {\bibfnamefont {A.}~\bibnamefont {Bansil}},\ }\href {\doibase
  10.1103/PhysRevB.88.165301} {\bibfield  {journal} {\bibinfo  {journal} {Phys.
  Rev. B}\ }\textbf {\bibinfo {volume} {88}},\ \bibinfo {pages} {165301}
  (\bibinfo {year} {2013})}\BibitemShut {NoStop}%
\bibitem [{\citenamefont {Yang}\ \emph {et~al.}(2012)\citenamefont {Yang},
  \citenamefont {Miao}, \citenamefont {Wang}, \citenamefont {Yao},
  \citenamefont {Zhu}, \citenamefont {Song}, \citenamefont {Wang},
  \citenamefont {Xu}, \citenamefont {Fedorov}, \citenamefont {Sun},
  \citenamefont {Zhang}, \citenamefont {Liu}, \citenamefont {Liu},
  \citenamefont {Qian}, \citenamefont {Gao},\ and\ \citenamefont {Jia}}]{bi-5}%
  \BibitemOpen
  \bibfield  {author} {\bibinfo {author} {\bibfnamefont {F.}~\bibnamefont
  {Yang}}, \bibinfo {author} {\bibfnamefont {L.}~\bibnamefont {Miao}}, \bibinfo
  {author} {\bibfnamefont {Z.~F.}\ \bibnamefont {Wang}}, \bibinfo {author}
  {\bibfnamefont {M.-Y.}\ \bibnamefont {Yao}}, \bibinfo {author} {\bibfnamefont
  {F.}~\bibnamefont {Zhu}}, \bibinfo {author} {\bibfnamefont {Y.~R.}\
  \bibnamefont {Song}}, \bibinfo {author} {\bibfnamefont {M.-X.}\ \bibnamefont
  {Wang}}, \bibinfo {author} {\bibfnamefont {J.-P.}\ \bibnamefont {Xu}},
  \bibinfo {author} {\bibfnamefont {A.~V.}\ \bibnamefont {Fedorov}}, \bibinfo
  {author} {\bibfnamefont {Z.}~\bibnamefont {Sun}}, \bibinfo {author}
  {\bibfnamefont {G.~B.}\ \bibnamefont {Zhang}}, \bibinfo {author}
  {\bibfnamefont {C.}~\bibnamefont {Liu}}, \bibinfo {author} {\bibfnamefont
  {F.}~\bibnamefont {Liu}}, \bibinfo {author} {\bibfnamefont {D.}~\bibnamefont
  {Qian}}, \bibinfo {author} {\bibfnamefont {C.~L.}\ \bibnamefont {Gao}}, \
  and\ \bibinfo {author} {\bibfnamefont {J.-F.}\ \bibnamefont {Jia}},\ }\href
  {\doibase 10.1103/PhysRevLett.109.016801} {\bibfield  {journal} {\bibinfo
  {journal} {Phys. Rev. Lett.}\ }\textbf {\bibinfo {volume} {109}},\ \bibinfo
  {pages} {016801} (\bibinfo {year} {2012})}\BibitemShut {NoStop}%
\bibitem [{\citenamefont {Hirahara}\ \emph {et~al.}(2011)\citenamefont
  {Hirahara}, \citenamefont {Bihlmayer}, \citenamefont {Sakamoto},
  \citenamefont {Yamada}, \citenamefont {Miyazaki}, \citenamefont {Kimura},
  \citenamefont {Bl\"ugel},\ and\ \citenamefont {Hasegawa}}]{bi-6}%
  \BibitemOpen
  \bibfield  {author} {\bibinfo {author} {\bibfnamefont {T.}~\bibnamefont
  {Hirahara}}, \bibinfo {author} {\bibfnamefont {G.}~\bibnamefont {Bihlmayer}},
  \bibinfo {author} {\bibfnamefont {Y.}~\bibnamefont {Sakamoto}}, \bibinfo
  {author} {\bibfnamefont {M.}~\bibnamefont {Yamada}}, \bibinfo {author}
  {\bibfnamefont {H.}~\bibnamefont {Miyazaki}}, \bibinfo {author}
  {\bibfnamefont {S.}~\bibnamefont {Kimura}}, \bibinfo {author} {\bibfnamefont
  {S.}~\bibnamefont {Bl\"ugel}}, \ and\ \bibinfo {author} {\bibfnamefont
  {S.}~\bibnamefont {Hasegawa}},\ }\href {\doibase
  10.1103/PhysRevLett.107.166801} {\bibfield  {journal} {\bibinfo  {journal}
  {Phys. Rev. Lett.}\ }\textbf {\bibinfo {volume} {107}},\ \bibinfo {pages}
  {166801} (\bibinfo {year} {2011})}\BibitemShut {NoStop}%
\bibitem [{\citenamefont {Agapito}\ \emph {et~al.}(2013)\citenamefont
  {Agapito}, \citenamefont {Ferretti}, \citenamefont {Calzolari}, \citenamefont
  {Curtarolo},\ and\ \citenamefont {Buongiorno~Nardelli}}]{PAO1}%
  \BibitemOpen
  \bibfield  {author} {\bibinfo {author} {\bibfnamefont {L.~A.}\ \bibnamefont
  {Agapito}}, \bibinfo {author} {\bibfnamefont {A.}~\bibnamefont {Ferretti}},
  \bibinfo {author} {\bibfnamefont {A.}~\bibnamefont {Calzolari}}, \bibinfo
  {author} {\bibfnamefont {S.}~\bibnamefont {Curtarolo}}, \ and\ \bibinfo
  {author} {\bibfnamefont {M.}~\bibnamefont {Buongiorno~Nardelli}},\ }\href
  {\doibase 10.1103/PhysRevB.88.165127} {\bibfield  {journal} {\bibinfo
  {journal} {Phys. Rev. B}\ }\textbf {\bibinfo {volume} {88}},\ \bibinfo
  {pages} {165127} (\bibinfo {year} {2013})}\BibitemShut {NoStop}%
\bibitem [{\citenamefont {Agapito}\ \emph {et~al.}(2015)\citenamefont
  {Agapito}, \citenamefont {Curtarolo},\ and\ \citenamefont
  {Buongiorno~Nardelli}}]{PAO2}%
  \BibitemOpen
  \bibfield  {author} {\bibinfo {author} {\bibfnamefont {L.~A.}\ \bibnamefont
  {Agapito}}, \bibinfo {author} {\bibfnamefont {S.}~\bibnamefont {Curtarolo}},
  \ and\ \bibinfo {author} {\bibfnamefont {M.}~\bibnamefont
  {Buongiorno~Nardelli}},\ }\href {\doibase 10.1103/PhysRevX.5.011006}
  {\bibfield  {journal} {\bibinfo  {journal} {Phys. Rev. X}\ }\textbf {\bibinfo
  {volume} {5}},\ \bibinfo {pages} {011006} (\bibinfo {year}
  {2015})}\BibitemShut {NoStop}%
\bibitem [{\citenamefont {Agapito}\ \emph
  {et~al.}(2016{\natexlab{a}})\citenamefont {Agapito}, \citenamefont {Fornari},
  \citenamefont {Ceresoli}, \citenamefont {Ferretti}, \citenamefont
  {Curtarolo},\ and\ \citenamefont {Nardelli}}]{PAO3}%
  \BibitemOpen
  \bibfield  {author} {\bibinfo {author} {\bibfnamefont {L.~A.}\ \bibnamefont
  {Agapito}}, \bibinfo {author} {\bibfnamefont {M.}~\bibnamefont {Fornari}},
  \bibinfo {author} {\bibfnamefont {D.}~\bibnamefont {Ceresoli}}, \bibinfo
  {author} {\bibfnamefont {A.}~\bibnamefont {Ferretti}}, \bibinfo {author}
  {\bibfnamefont {S.}~\bibnamefont {Curtarolo}}, \ and\ \bibinfo {author}
  {\bibfnamefont {M.~B.}\ \bibnamefont {Nardelli}},\ }\href {\doibase
  10.1103/PhysRevB.93.125137} {\bibfield  {journal} {\bibinfo  {journal} {Phys.
  Rev. B}\ }\textbf {\bibinfo {volume} {93}},\ \bibinfo {pages} {125137}
  (\bibinfo {year} {2016}{\natexlab{a}})}\BibitemShut {NoStop}%
\bibitem [{\citenamefont {Agapito}\ \emph
  {et~al.}(2016{\natexlab{b}})\citenamefont {Agapito}, \citenamefont
  {Ismail-Beigi}, \citenamefont {Curtarolo}, \citenamefont {Fornari},\ and\
  \citenamefont {Nardelli}}]{PAO4}%
  \BibitemOpen
  \bibfield  {author} {\bibinfo {author} {\bibfnamefont {L.~A.}\ \bibnamefont
  {Agapito}}, \bibinfo {author} {\bibfnamefont {S.}~\bibnamefont
  {Ismail-Beigi}}, \bibinfo {author} {\bibfnamefont {S.}~\bibnamefont
  {Curtarolo}}, \bibinfo {author} {\bibfnamefont {M.}~\bibnamefont {Fornari}},
  \ and\ \bibinfo {author} {\bibfnamefont {M.~B.}\ \bibnamefont {Nardelli}},\
  }\href {\doibase 10.1103/PhysRevB.93.035104} {\bibfield  {journal} {\bibinfo
  {journal} {Phys. Rev. B}\ }\textbf {\bibinfo {volume} {93}},\ \bibinfo
  {pages} {035104} (\bibinfo {year} {2016}{\natexlab{b}})}\BibitemShut
  {NoStop}%
\bibitem [{\citenamefont {Nardelli}\ \emph {et~al.}(2018)\citenamefont
  {Nardelli}, \citenamefont {Cerasoli}, \citenamefont {Costa}, \citenamefont
  {Curtarolo}, \citenamefont {Gennaro}, \citenamefont {Fornari}, \citenamefont
  {Liyanage}, \citenamefont {Supka},\ and\ \citenamefont {Wang}}]{PAO5}%
  \BibitemOpen
  \bibfield  {author} {\bibinfo {author} {\bibfnamefont {M.~B.}\ \bibnamefont
  {Nardelli}}, \bibinfo {author} {\bibfnamefont {F.~T.}\ \bibnamefont
  {Cerasoli}}, \bibinfo {author} {\bibfnamefont {M.}~\bibnamefont {Costa}},
  \bibinfo {author} {\bibfnamefont {S.}~\bibnamefont {Curtarolo}}, \bibinfo
  {author} {\bibfnamefont {R.~D.}\ \bibnamefont {Gennaro}}, \bibinfo {author}
  {\bibfnamefont {M.}~\bibnamefont {Fornari}}, \bibinfo {author} {\bibfnamefont
  {L.}~\bibnamefont {Liyanage}}, \bibinfo {author} {\bibfnamefont {A.~R.}\
  \bibnamefont {Supka}}, \ and\ \bibinfo {author} {\bibfnamefont
  {H.}~\bibnamefont {Wang}},\ }\href {\doibase
  https://doi.org/10.1016/j.commatsci.2017.11.034} {\bibfield  {journal}
  {\bibinfo  {journal} {Computational Materials Science}\ }\textbf {\bibinfo
  {volume} {143}},\ \bibinfo {pages} {462 } (\bibinfo {year}
  {2018})}\BibitemShut {NoStop}%
\bibitem [{SM()}]{SM}%
  \BibitemOpen
  \href@noop {} {\bibinfo  {journal} {{Supplemental Material}}\ }\BibitemShut
  {NoStop}%
\bibitem [{\citenamefont {Costa}\ \emph {et~al.}(2010)\citenamefont {Costa},
  \citenamefont {Muniz}, \citenamefont {Lounis}, \citenamefont {Klautau},\ and\
  \citenamefont {Mills}}]{antc:2010:SOCMethod}%
  \BibitemOpen
\bibfield  {journal} {  }\bibfield  {author} {\bibinfo {author} {\bibfnamefont
  {A.~T.}\ \bibnamefont {Costa}}, \bibinfo {author} {\bibfnamefont {R.~B.}\
  \bibnamefont {Muniz}}, \bibinfo {author} {\bibfnamefont {S.}~\bibnamefont
  {Lounis}}, \bibinfo {author} {\bibfnamefont {A.~B.}\ \bibnamefont {Klautau}},
  \ and\ \bibinfo {author} {\bibfnamefont {D.~L.}\ \bibnamefont {Mills}},\
  }\href {\doibase 10.1103/PhysRevB.82.014428} {\bibfield  {journal} {\bibinfo
  {journal} {Phys. Rev. B}\ }\textbf {\bibinfo {volume} {82}},\ \bibinfo
  {pages} {014428} (\bibinfo {year} {2010})}\BibitemShut {NoStop}%
\bibitem [{\citenamefont {Guimar\~aes}\ \emph {et~al.}(2015)\citenamefont
  {Guimar\~aes}, \citenamefont {Lounis}, \citenamefont {Costa},\ and\
  \citenamefont {Muniz}}]{FilipePRB2015}%
  \BibitemOpen
  \bibfield  {author} {\bibinfo {author} {\bibfnamefont {F.~S.~M.}\
  \bibnamefont {Guimar\~aes}}, \bibinfo {author} {\bibfnamefont
  {S.}~\bibnamefont {Lounis}}, \bibinfo {author} {\bibfnamefont {A.~T.}\
  \bibnamefont {Costa}}, \ and\ \bibinfo {author} {\bibfnamefont {R.~B.}\
  \bibnamefont {Muniz}},\ }\href {\doibase 10.1103/PhysRevB.92.220410}
  {\bibfield  {journal} {\bibinfo  {journal} {Phys. Rev. B}\ }\textbf {\bibinfo
  {volume} {92}},\ \bibinfo {pages} {220410} (\bibinfo {year}
  {2015})}\BibitemShut {NoStop}%
\bibitem [{\citenamefont {Zhu}\ \emph {et~al.}(2011)\citenamefont {Zhu},
  \citenamefont {Yao}, \citenamefont {Zhang},\ and\ \citenamefont
  {Chang}}]{rkky1}%
  \BibitemOpen
  \bibfield  {author} {\bibinfo {author} {\bibfnamefont {J.-J.}\ \bibnamefont
  {Zhu}}, \bibinfo {author} {\bibfnamefont {D.-X.}\ \bibnamefont {Yao}},
  \bibinfo {author} {\bibfnamefont {S.-C.}\ \bibnamefont {Zhang}}, \ and\
  \bibinfo {author} {\bibfnamefont {K.}~\bibnamefont {Chang}},\ }\href
  {\doibase 10.1103/PhysRevLett.106.097201} {\bibfield  {journal} {\bibinfo
  {journal} {Phys. Rev. Lett.}\ }\textbf {\bibinfo {volume} {106}},\ \bibinfo
  {pages} {097201} (\bibinfo {year} {2011})}\BibitemShut {NoStop}%
\bibitem [{\citenamefont {Abanin}\ and\ \citenamefont {Pesin}(2011)}]{rkky2}%
  \BibitemOpen
  \bibfield  {author} {\bibinfo {author} {\bibfnamefont {D.~A.}\ \bibnamefont
  {Abanin}}\ and\ \bibinfo {author} {\bibfnamefont {D.~A.}\ \bibnamefont
  {Pesin}},\ }\href {\doibase 10.1103/PhysRevLett.106.136802} {\bibfield
  {journal} {\bibinfo  {journal} {Phys. Rev. Lett.}\ }\textbf {\bibinfo
  {volume} {106}},\ \bibinfo {pages} {136802} (\bibinfo {year}
  {2011})}\BibitemShut {NoStop}%
\bibitem [{Note1()}]{Note1}%
  \BibitemOpen
  \bibinfo {note} {In linear response theory the system is assumed to be
  arbitrarily close to equilibrium at all times. This means that the excitation
  energy is determined by the ground state electronic structure, and may differ
  from the MAE calculated via total energy differences.}\BibitemShut {Stop}%
\bibitem [{\citenamefont {Costa}\ \emph {et~al.}(2008)\citenamefont {Costa},
  \citenamefont {Muniz}, \citenamefont {Ferreira},\ and\ \citenamefont
  {Mills}}]{antc:2008:DynamicCouplingToyModel}%
  \BibitemOpen
  \bibfield  {author} {\bibinfo {author} {\bibfnamefont {A.~T.}\ \bibnamefont
  {Costa}}, \bibinfo {author} {\bibfnamefont {R.~B.}\ \bibnamefont {Muniz}},
  \bibinfo {author} {\bibfnamefont {M.~S.}\ \bibnamefont {Ferreira}}, \ and\
  \bibinfo {author} {\bibfnamefont {D.~L.}\ \bibnamefont {Mills}},\ }\href
  {\doibase 10.1103/PhysRevB.78.214403} {\bibfield  {journal} {\bibinfo
  {journal} {Phys. Rev. B}\ }\textbf {\bibinfo {volume} {78}},\ \bibinfo
  {pages} {214403} (\bibinfo {year} {2008})}\BibitemShut {NoStop}%
\end{thebibliography}%

\end{document}